# DETERMINATION OF THE HUBBLE CONSTANT USING CEPHEIDS

MOHAMED ABDEL-SABOUR [1], MOHAMED EBRAHIM NOUH [1], ISSA ALI ISSA [1], MOHAMED SALEH EL-NAWAWY [2,3], AYMAN KORDI [3], ZAKI ALMOSTAFA [3,4], AHMAD ESSAM EL-SAID [1], GAMAL BAKR ALI [1],

[1] *National Research Institute of Astronomy and Geophysics (NRIAG)*
*11421-Helwan, Cairo, Egypt*
E-mails: `msabour@nriag.sci.eg`, `sabour2000@hotmail.com`,
`abdo_nouh@hotmail.com`, `essam60@yahoo.com`, `gamalali@lycos.com`

[2] *Astronomy Department, Faculty of Science, Cairo University*
*Cairo, Egypt*
E-mail: `msnawawy@yahoo.com`

[3] *Physics and Astronomy Department, Faculty of Science, King Saud University*
*Riyadh, Saudi Arabia*
E-mail: `aymankurdi@yahoo.com`

[4] *King Abdelaziz City for Science Technology*
*Riyadh, Saudi Arabia*
E-mail: `zalmostafa@kacst.edu.sa`

*Abstract.* This paper introduces a statistical treatment to use Cepheid variable stars as distance indicators. The expansion rate of the Universe is also studied here through deriving the value of the Hubble constant $H_0$. A Gaussian function approximation is proposed to fit the absolute magnitude and period of Cepheid variables in our galaxy. The calculations are carried out on samples of Cepheids observed in 23 galaxies to derive the distance modulus (DM) of these galaxies based on the frequency distributions of their periods and intrinsic apparent magnitudes.

The DM is the difference between the apparent magnitude for extragalactic Cepheids and the absolute magnitude of the galactic Cepheids at maximum number. It is calculated by using the comparison of the period distribution of Cepheids in our galaxy and in other galaxies.

This method is preferred due to its simplicity to use and its efficiency in providing reliable DM. A linear fit with correlation coefficient of 99.68% has been found between the published distance modulus and that computed one in the present work. From the present sample, a value of $H_0$ in the range of 66 to $80 \pm 5$ km s$^{-1}$ Mpc$^{-1}$ is determined.

The present procedure of computation and its accuracy are confirmed by the high correlation found between our computed DM and that published in the literature.

## 1. INTRODUCTION

There are various methods that have been proposed and applied to determine the distances of celestial objects as premier data aimed at understanding the physics of the universe. Among the most important of these methods is the consideration of Cepheid variables as standard candles. As summarized by Freedman et al. (2001) and Teerikorpi and Paturel (2002), Cepheids represent the critical initial step on the cosmic distance ladder. Hubble was able to use his observational data for Cepheids in galaxies to deduce a law, which bears his name. This law states that the more remote the galaxy is from the Milky Way, the faster its recession is:

$$V = H_0 d, \qquad (1)$$

where $H_0$ has a value between 50 and 100 km s$^{-1}$Mpc$^{-1}$. To study the Hubble constant we need to focus on a reliable distance indicators. Cepheid variables are one of the most useful tracers. To find out the distance of an extra-galaxy, there are many complicated points regarding the PL relations. Many efforts have been carried out to explore the PL relation and some others relations like PLC, PC, and PA relations.

In Section 2 a review of the studies of Cepheids as distance indicator is given. Section 3 surveys the previous studies done on the Hubble constant.

## 2. CEPHEID VARIABLES AS DISTANCE INDICATORS

Since Henrietta Leavitt until now, the period-luminosity relation (PL) for Cepheids is the most valuable item in the tool kit of extragalactic astronomers. This PL relation remains one of the most powerful tools that astronomers have at their disposal for measuring the distance to nearby galaxies, since, beyond the Local Group, the reddening is a very hard problem we cannot solve exactly. There are two main sources of the interstellar reddening: reddening in the star's parent galaxy and dust in our own Galactic foreground. For this reason, when we need to determine the absolute magnitude we use multi-wavelength observations to deal with the reddening problem. Madore and Freedman (1991) suggested that the reddening problem should be solved before any empirical determination of the coefficients in the PLC relation. The theoretical calculations predict a finite width to the instability strip (with temperature/color being the control parameters). In addition, metallicity is a quantity that is known to be different from galaxy to galaxy. Calibration of the PL relation of Cepheids on the basis of the extragalactic distance scale is still uncertain at $\Delta M = \pm 0.1$ mag.

The determination of accurate reddening for Cepheids has traditionally followed three distinct routes:

(i) from field reddening of specific objects (cluster and association Cepheids),

based on the analysis of photometric data for early type stars sharing the same lines of sight;

(ii) from observations mainly of bright Cepheids involving photometric parameters designed to be independent of interstellar reddening;

(iii) from using standard reddening laws and a calibrated intrinsic color scheme to de-redden photometric observations for large samples of Cepheids.

Methods (i) and (ii) are the most reliable means of establishing reddening for individual Cepheids, since method (iii) often involves the use of period-color relation, which does not necessarily account for the intrinsic spread of Cepheid instability strip.

There also are difficulties associated with measuring Cepheid distances:

*First*, since Cepheids are young stars, they are found in regions where there is dust scattering, absorption, and reddening. Extinction is systematic, and its effects must be either removed by multicolor data or minimized by observing at long wavelengths, or both.

*Second*, the dependence of the PL relation on metallicity has been very difficult to quantify.

*Third*, an accurate geometric calibration of the PL relation, for any given metallicity, has not been established yet (Freedman et al. 2001).

Regarding Cepheids, we follow the same criteria of minimizing factors that may affect the distance determination. The magnitude and period distribution functions of Galactic Cepheids are studied in both our galaxy and beyond. Cepheids in the Milky Way and the LMC do follow different PL relations, and hence the Cepheid PL relation is not universal (Ngeow and Kanbur 2004). The Galactic PL relation is steeper than the LMC counterparts (Tammann et al. 2003a; Fouque et al, 2003; Storm et al. 2004). For this reason we propose this statistical approach to estimate the distance modulus for galaxies in the Local Group.

### 3. HUBBLE CONSTANT

Since Hubble's discovery of the redshift – apparent magnitude relation for galaxies, the value of the slope of the expansion law has been in contention. In the 1960's there were two camps, one claiming a value around 50 and the other claiming around $H_0 = 100 \, \text{km s}^{-1} \text{Mpc}^{-1}$. In the last 20 years, much progress has been made and estimates now range between 60 and 75 km s$^{-1}$Mpc$^{-1}$.

A huge improvement in the calculation of $H_0$ is given by a project using HST. The aim of this project was to derive a value for the expansion rate of the Universe, the Hubble constant, to an accuracy of 10% (see Freedman et al. 1994; Kennicutt et al. 1995; Mould et al. 1995; Madore et al. 1998). The value given by Freedman et al. (2001) was revised to lying between 72 km s$^{-1}$Mpc$^{-1}$ and $76 \pm 8$ km s$^{-1}$Mpc$^{-1}$. As mentioned by Floor et al. (2007), these results are insensitive to Cepheid metallicity corrections. Sandage et al.(2006), in the final result about the Key project of Hubble Space Telescope, after adopting metallicity for the Cepheid distance, they found $H_0 = 62.3 \pm 1.3$ km s$^{-1}$Mpc$^{-1}$. They used not only Cepheid variables, but also the apparent magnitudes of 62 normal SNe Ia with $3{,}000 < v_{cmb} < 20{,}000$ km/s.

We aimed at understanding the reasons for which independent analyses about the same data give values which are discrepant by twice the quoted systematic errors. Probably the fairest and most up-to-date analysis is achieved by comparing the result of Riess et al. (2005) (R05), who found $H_0 = 73 \pm 4$ km s$^{-1}$Mpc$^{-1}$, and the one of Sandage et al. (2006) (S06) who found $H_0 = 62.3 \pm 1.3$ km s$^{-1}$Mpc$^{-1}$. Both appear quoting a systematic error of 5 km s$^{-1}$Mpc$^{-1}$.

The R05 analysis is based on four SNe Ia: 1994ae in NGC 3370, 1998aq in NGC 3982, 1990N in NGC 4639 and 1981B in NGC 4536. The S06 analysis included eight other calibrators; there is not an issue as S06 find $H_0 = 63.3 \pm 1.9$ km s$^{-1}$Mpc$^{-1}$ from these four calibrators separately, still a 15% difference. The comparison of Table 13 of R05 and Table 1 of S06 reveals part of the problem; the distances of the four calibrators are generally discrepant in the two analyses, with S06 having the higher value. In the best case, SN 1994ae in NGC 3370, the discrepancy is only 0.08 mag. In the worst case (SN1990N in NGC4639) R05 quotes a distance modulus $\mu_0 = 31.74$, whereas the value obtained by S06 is 32.20, corresponding to a 20–25% difference in the inferred distance and hence in $H_0$. The quoted distance modulus, $\mu_0$ is formed by a combination of observations at two optical bands, $V$ and $I$, although the coefficients differ slightly between different authors. The purpose of the combination is to eliminate differential effects due to reddening, which does exactly provide the reddening law. The two groups R05 and S06 have used different corrections to the metallicity. This leads to differences in their results of the distances of galaxies. The group R05 have used a global correction $\Delta\mu = 0.24\Delta[O/H]$ given by Sakai et al. (2004) ($\Delta[O/H]$ represents the metallicity of the observed Cepheids minus the metallicity of the LMC), whereas S06's correction merges by interpolation between LMC and the Galactic PL relations which gives (Saha et al. 2006) $\Delta\mu = 1.6(\log P - 0.933)[O/H]$. The calculation given by R05 seems to be in favor for us. Hence it will be used as a matter of comparison with our results for the Cepheid distribution in the Local Group.

### 4. THE FORMS OF PLR AND PLC RELATIONS

Cepheids have long been considered as the most trustful extragalactic distance indicator (Sandage and Tammann 1969; Feast and Walker 1987). Many authors use the PL relation (Stothers 1983), others use the PLC relation (Laney and Stobie 1986), whereas a third group believes in the PLA relation (Sandage et al. 2004). Caldwell (1985) took the metallicity into account, and some others neglect metallicity. Interstellar extinction still remains a very serious problem in the cosmological distance ladder when using standard candles procedure. The main difficulty arises in estimating color indices, color excesses and not least the assumed value of the total to selective absorption ratio, $R = A_V / E(B-V)$ (Turner 1976). In some locations in the sky, some authors assume $R$ equal to 3.1, others assume a value of 3.3 and a third adopts $R$ as 3.0. In some areas of the Galaxy it reaches a value of 6, and even 7. Whenever facing interstellar extinction, two problems arise, internal extinction depending on the distance travelled by the light of the standard candle from its position in the parent galaxy and its location. Following is the extinction while penetrating our galaxy to the observer. This depends also on where it penetrates our Galaxy. Still debate is the extinction of light at the poles of the galaxy. Is it negligible according to Sandage (1984) or does it follow the $\csc b$ law as claimed by de Vaucouleurs et al (1976) and de Vaucouleurs (1983a, b).

For Cepheids, Novae and other standard candles used as distance indicators, we want to generalize the de Vaucouleurs say in (1978), "the unending discussions, revisions, and re-discussions of the PL, PLC and PLA relations make the point clear, it is evolution that makes these differences". Accordingly, all standard candles have to be considered of the same weight whenever estimating distances to galaxies. To know how errors propagate, let us consider the following cases for Cepheids. The PLC and PC relations have been given by Robinson (1988) as

$$M_V = -n - m \log P + q(B-V), \qquad (2)$$

where

$$(B-V) = s \log P + t. \qquad (3)$$

The coefficients of these relations are subjected to a large scattering, with average amounts of 3.7% for the value of $n$, 3.5% for $m$, and 9.3% for $q$.

Tammann et al. (2003), using modern data, found that the values of $s$ and $t$ for Galactic Classical Cepheids are $0.366 \pm 0.015$ and $0.361 \pm 0.013$, respectively. The relative scattering in $M_V$ is due to the scattering in the values of $n$, $m$ and $q$ among different authors, which can be given by

$$\frac{\sigma M_V}{M_V} = \pm \sqrt{\left(\frac{\sigma n}{n}\right)^2 + \left(\frac{\sigma m}{m}\right)^2 + \left(\frac{\sigma q}{q}\right)^2}, \qquad (4)$$

assuming that $n$, $m$, $q$ are independent factors. The scattering in $M_V$ due to the scattering of $n$, $m$, $q$ values reached the high percentage. We want to remind that this value is due to different estimations in the values of the coefficients of the PLC and PC relation from different authors. The values given for *m* in Table 3.2 provided by Robinson (1988) ranges between $m = 2.79$ (Fernie 1967 from theory) and $m = 4.11$ (Van den Bergh 1977). Which of them can we use?

The average distance modulus of M31 as deduced from Table 3.4 given by Robinson (1988) is 24.32 in a range of $24.09 \leq \mu \leq 24.60$ as deduced from Cepheids, while the average is 24.04 as deduced using novae (Robinson 1988, Table 3.6). The average difference is 0.28 mag. The first average makes a distance of 731 kpc while the second makes 643 kpc with a difference of 88 kpc. If we assume that M31 is as big as our Galaxy with a major axis of 100,000 LY or 30 kpc, then the error in estimating the distance to M31 is nearly 3 times as big as its major axis as a whole. The scattering among different authors' values of the $(B-V)$ of equation (3) is 43%.

Among different distance indicators, we find a systematic difference of the order of $1\,\text{mag}$ between distances evaluated by Tulley-Fisher method and those derived by supernovae (SN) (Robinson 1988). It has to be mentioned that the scattering in observations in one of the test samples using the method suggested in the present work, as it was mentioned by the authors, ranges for M31 visual magnitudes (see Baade and Swope 1965) between $0.02\,\text{mag}$ to $0.2\,\text{mag}$, although it is always mentioned that this is one of the most accurate catalogues. While considering a distance indicator, it is required to minimize as much as possible the factors affecting distance moduli, as tried by Issa (1980). The radius distribution function of some objects like dark clouds (Issa 1980), HII-regions (Issa 1985), globular clusters (Issa 1989), were used to determine distances to galaxies.

### 5. CALIBRATION FOR GALACTIC AND EXTRAGALACTIC CEPHEIDS

Many samples of Galactic Cepheids have been selected as observational materials for the present work. The first sample contains 451 galactic Cepheids taken from the Catalogue of Berdnikov et al. (2000a, b). The second sample includes 324 Galactic Cepheids that have been taken from Tammann et al. (2003b), and the third sample is taken from David Dunlap Observatory (DDO) and can be found in the site `http://www.astro.utoronto.ca/DDO/research/cepheids/table_physical.html`.

The method proposed in the present work is based on modeling the distribution of each of the Cepheids absolute magnitudes and periods using the distribution function described in Subsection 3.1. The Extragalactic Cepheid observations were undertaken from the data of Hubble Space Telescope, which can be found in the archive at the site `http://www.ipac.caltech.edu/H0kp` and given by Wolfgang et al. (1993).

## 6. BASIC CONCEPT

In the present work we assume Gaussian fits for both the absolute magnitude of Cepheids in our galaxy and their apparent magnitudes.

Our approach has two stages:

*First*, using Galactic Cepheids, we determine the absolute magnitude at maximum number.

*Second*, the distribution function of the apparent magnitudes of Cepheids in galaxies under study is fitted by Gaussian function to deduce the magnitude at maximum number. The period distribution is fitted with a Gaussian, as well, and the absolute magnitude corresponding to the maximum period was also deduced from the PL relation and also used to deduce the distance of the galaxy.

The division into magnitude intervals follows from the statistical relation $K \leq 5 \log N$, where $K$ is the number of magnitude intervals and $N$ is the number of stars. Such method of sampling can exhibit the global characteristics of the sample. The natural logarithm of the number is plotted against the mid-magnitude of each interval. In all cases the Gaussian fits are shown in Figs. 1 and 2.

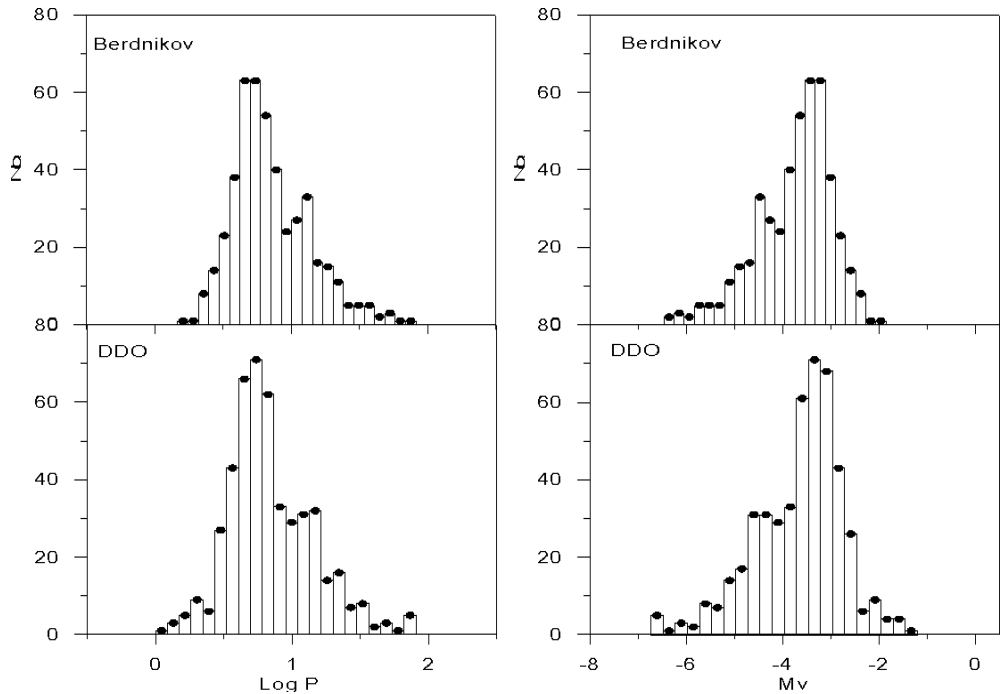

Fig. 1 – Histograms for the period and the absolute magnitude distributions for Galactic Cepheids from Berdnikov et al. (2000a, b) and DDO Catalogue, respectively.

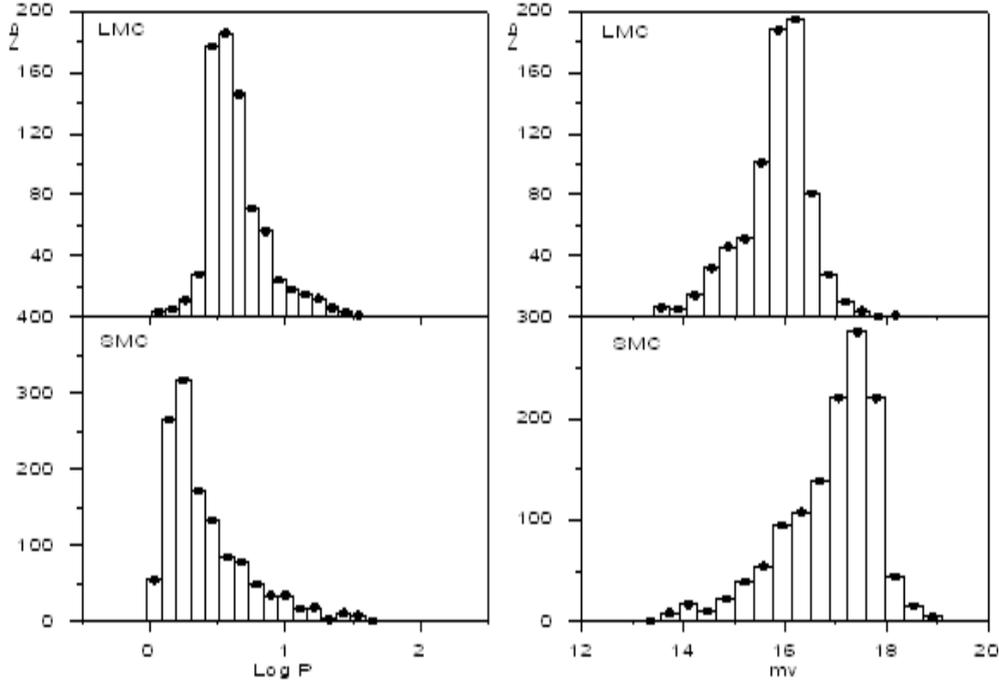

Fig. 2 – Histograms for Period and apparent magnitude distribution for LMC and SMC.

For Cepheids of our Galaxy taken from the Berdnikov et al. (2000a, b) and Tammann et al. (2003a, b) catalogues, we have calculated the absolute magnitude using the relation, $M_V = -1.4 - 2.76 \log P$, which is given by Madore and Freedman (1991). But the PL relation, for the data taken from the DDO observations, is calculated using the Fernie (1992) relation, $M_V = -1.203 - 2.902 \log P$.

Applying the Gaussian function to the sample of Berdnikov et al. (2000a, b) and of Tammann et al. (2003b) gives respectively

$$\ln N = 65.198 \exp\left[-\frac{(M - 3.99)^2}{2(0.941)^2}\right], \qquad (5)$$

$$\ln N = 16.315 \exp\left[-\frac{(M - 4.193)^2}{2(0.956)^2}\right]. \qquad (6)$$

The mean absolute visual magnitude at maximum number of Galactic Cepheids for the first sample is $-3.99$ mag, while that for the second sample is $-4.2$ mag. The standard deviations around these averages are $\pm 0.941$ and $\pm 0.956$, respectively.

The Gaussian distribution for Galactic and Extragalactic Cepheids may be written in the following form:

*Galactic Cepheids*:

$$N = a \exp\left[-\frac{(M-b)^2}{2\sigma^2}\right], \tag{7a}$$

hence

$$\ln N = \ln a - \frac{(M-b)^2}{2\sigma^2}, \tag{7b}$$

or

$$S = C_1 - \frac{(M-\overline{M})^2}{C_2}, \tag{7c}$$

where $S = \ln N$, $C_1 = \ln a$, $C_2 = 2\sigma^2$.

Therefore

$$(M - \overline{M}) = \sqrt{C_1 C_2 - C_2 S}, \tag{8}$$

where $N$ is the number of stars corresponding to absolute magnitude at maximum number $\overline{M}$.

*Extragalactic Cepheids*:

For any other galaxy whose apparent magnitude $m_V$ is used, we get a similar relation of the form

$$n = d \exp\left[-\frac{(m-b)^2}{2\sigma'^2}\right], \tag{9a}$$

hence

$$\ln n = \ln d - \frac{(m-b)^2}{2\sigma'^2}, \tag{9b}$$

or

$$s = c_3 - \frac{(m-\overline{m})^2}{c_4}, \quad (9c)$$

where $s = \ln n$, $c_3 = \ln d$, $c_4 = 2\sigma'^2$.

Therefore,

$$(m - \overline{m}) = \sqrt{c_3 c_4 - c_4 s}, \quad (10)$$

where $n$ is the number of stars corresponding to the maximum apparent magnitude ($m$).

Subtracting (8) from (10) we get,

$$(m - \overline{m}) - (M - \overline{M}) = \sqrt{c_3 c_4 - c_4 s} - \sqrt{C_1 C_2 - C_2 S}. \quad (11)$$

From equations (7) and (9), and knowing the Gaussian parameters, at maximum number we have, $M = \overline{M}$, which gives $S = C_1$, and $m = \overline{m}$, which gives $s = c_3$, respectively, therefore we have from (11), $(m - \overline{m}) - (M - \overline{M}) = 0$, following that

$$(m - \overline{m}) = (M - \overline{M}) \Rightarrow m - M = \overline{m} - \overline{M}. \quad (12)$$

Equation (12) indicates that the Gaussian distribution predicts that the distance modulus is equal to its value at maximum number.

## 7. RESULTS AND DISCUSSION

The new approach was applied to samples of Cepheids in 23 galaxies to derive the distance modulus (DM) of these galaxies based on the assumption that absolute magnitudes, apparent magnitudes, and the periods of Cepheids do follow Gaussian distribution. This assumption is not far from being correct since the mechanism of Cepheid variability, the physical process that leads to the Cepheid phenomenon, is universal The results obtained are given in Table 1, where the columns contain: the name of the host galaxy, the number of Cepheids, the published DM, the present DM derived (with their errors), the average of the present findings, the sources of the published DM and the data employed, respectively. The data have been extracted from different sources. In particular there are eight galaxies have been collected from different sources (viz. NGCs 224, 925, 1365, 1425, 2090, 3621, 4321 and 5457). For these galaxies, small differences have been found between the derived DMs. These differences can be attributed to the different accuracies in the data. DM derived from either the period or the magnitude are very close; except for SMC and LMC, hence their mean values may be considered to compare the present DM derived values with those found in the literature.

*Table 1*

DM for Cepheids in 23 galaxies

| Name of galaxy | No. | Distance modulus (DM) ($m - M$) | | | | References | |
|---|---|---|---|---|---|---|---|
| | | $\mu_V$ Published | Present values (DM) | | | Distance Publ. | Data |
| | | | DM (Period) | DM (Magnitude) | Mean DM | | |
| NGC224(M31) | 130 | 24.47±0.05 | 24.80±0.04 | 24.95±0.08 | 24.91±0.05 | 8 | 3 |
| | 38 | | 25.39±0.05 | 25.16±0.22 | 25.27±0.15 | 5 | 1 |
| NGC598(M33) | 251 | 24.56±0.10 | 25.06±0.04 | 25.03±0.25 | 25.06±0.14 | 5 | 3 |
| NGC925 | 61 | 29.80±0.04 | 29.62±0.03 | 29.89±0.15 | 29.75±0.14 | 5 | 1 |
| | 71 | | 30.01±0.03 | 29.61±0.14 | 29.81±0.10 | 5 | 2 |
| NGC1365 | 40 | 31.18±0.05 | 30.86±0.02 | 30.61±0.04 | 30.73±0.03 | 5 | 1 |
| | 52 | | 30.86±0.02 | 30.66±0.06 | 30.76±0.04 | 5 | 2 |
| NGC1425 | 29 | 31.60±0.05 | 31.09±0.15 | 31.12±0.05 | 31.09±0.07 | 5 | 1 |
| | 29 | | 31.08±0.22 | 31.11±0.10 | 31.09±0.14 | 5 | 2 |
| NGC2090 | 31 | 30.29±0.04 | 30.10±0.09 | 30.12±0.18 | 30.12±0.13 | 5 | 1 |
| | 34 | | 30.03±0.05 | 29.88±0.10 | 29.94±0.07 | 5 | 2 |
| NGC2541 | 28 | 30.25±0.05 | 30.15±0.07 | 30.16±0.10 | 30.15±0.08 | 5 | 2 |
| NGC3031(M81) | 31 | 27.75±0.08 | 27.62±0.08 | 27.57±0.40 | 27.59±0.20 | 5 | 2 |
| NGC3198 | 77 | 30.68±0.08 | 30.80±0.04 | 30.31±0.17 | 30.55±0.11 | 5 | 1 |
| NGC3319 | 33 | 30.64±0.09 | 30.44±0.22 | 30.23±0.15 | 30.33±0.17 | 5 | 2 |
| NGC3351 | 49 | 29.85±0.09 | 30.25±0.21 | 30.12±0.08 | 30.18±0.12 | 5 | 2 |
| NGC3621 | 46 | 29.08±0.06 | 29.37±0.02 | 29.21±0.10 | 29.29±0.06 | 5 | 1 |
| | 69 | | 29.42±0.02 | 29.43±0.52 | 29.42±0.02 | 5 | 2 |
| NGC3627 | 105 | 29.86±0.08 | 29.90±0.02 | 29.68±0.04 | 29.79±0.03 | 5 | 1 |
| NGC4321(M100) | 42 | 30.78±0.07 | 30.70±0.03 | 30.57±0.11 | 30.63±0.07 | 5 | 1 |
| | 52 | | 30.54±0.03 | 30.64±0.50 | 30.58±0.22 | 5 | 2 |
| NGC4496A | 142 | 30.81±0.03 | 30.83±0.03 | 30.30±0.02 | 30.56±0.03 | 5 | 1 |
| NGC4535 | 50 | 30.85±0.05 | 30.48±0.02 | 30.42±0.38 | 30.44±0.20 | 5 | 2 |
| NGC4536 | 75 | 30.80±0.04 | 30.85±0.03 | 30.35±0.02 | 30.60±0.03 | 5 | 1 |
| NGC4548(M91) | 24 | 30.88±0.05 | 30.77±0.55 | 30.52±0.13 | 30.64±0.03 | 5 | 2 |
| NGC5457 (M101) | 31 | 29.13±0.11 | 28.87±0.06 | 28.46±0.12 | 28.66±0.09 | 5 | 1 |
| | 30 | | 28.85±0.09 | 28.73±0.03 | 28.79±0.06 | 5 | 2 |
| SMC | 1287 | 18.73±0.03 | 19.84±0.02 | 19.56±0.06 | 19.70±0.04 | 6 | 4 |
| LMC | 762 | 18.22±0.05 | 19.48±0.03 | 19.12±0.03 | 19.30±0.03 | 6 | 4 |
| IC4182 | 52 | 28.28±0.06 | 28.48±0.08 | 28.05±0.10 | 28.26±0.07 | 5 | 1 |
| IC1613 | | 24.19±0.15 | 24.21±0.19 | 24.29±0.22 | 24.20±0.20 | 5 | 7 |

The numbers in the last two columns correspond to: (1) Gieren et al. (1993); (2) HST (`http://www.ipac.caltech.edu/H0kp/H0keyProj.html`); (3) Magnier et al. (1997); (4) OGLE II; (5) Freedman et al. (2001); (6) Udalski et al. (1999); (7) Antonello et al. (2006); (8) Stanek and Garnavich (1998).

Nevertheless, the DMs derived from the period distribution are generally associated with less error in comparison with those based on the magnitude distribution. This is due to the known fact that the Cepheid period can be determined with higher precision than the magnitude resulting from de-reddening. The mean difference between these two DMs is 0.164 ($s.d. = \pm 0.201$) with a range from –0.269 to 0.526 if the DM of the LMC is excluded. The linear correlation coefficient is 99.669%. This may imply that the grouping and the Gaussian fit processes have been done properly. The average deviation between the published DM and those derived in the current work for the period, magnitude and average are 0.071 ($s.d. = \pm 0.258$), 0.259 ($s.d. = \pm 0.311$) and 0.165 ($s.d. = \pm 0.260$), respectively. The largest range of these deviations is associated with the DM obtained from the magnitude distribution. A linear fit, within 2 sigma, has been performed to correlate the published DM and the present mean values, leading to the following relation:

$$DM_{Pub.} = 1.009(\pm 0.010) DM_{ave} - 0.010(\pm 0.305),$$

where the probable errors are given in parentheses. The root mean square of the residuals is 0.254 and the linear correlation coefficient is 99.687%. It is evident that the zero-point term is very small in comparison with its error and the slope is almost unity. This relation is illustrated in Fig. 3 (A).

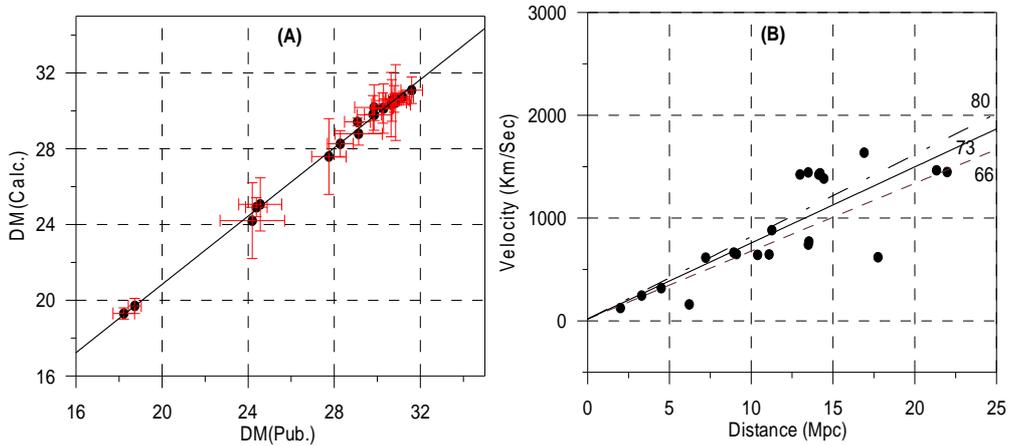

Fig. 3 – (A) The correlation between the published DM and the present mean value, and (B) velocity vs. distance for galaxies with Cepheid distances.

From this relation, it can be shown that the DMs obtained in the present study are in very good agreement with the published results. It has to be noticed that the DM ranges from 18.22 to 31.60 magnitudes. Regarding DM for SMC and LMC, where there is difference between our results and those published, we find that our result is near the published using RR Lyrae as given by Udalski et al (1999). They got $18.94 \pm 0.04$ and $19.44 \pm 0.03$ for the DM of LMC and SMC, respectively. Also we can attribute the large difference, appeared in our calculated distances of Magellanic Clouds in comparison to the referred published results, to the fact that we have considered the Cepheids without their separation into two modes as given in the catalogue given in the project OGLE II (Udalski et al. 1999). They divided Cepheids into two groups, fundamental mode (FU) and first overtone mode (FO), where FO on the average are brighter than FU. The calculated distances of the considered galaxies belonging to the Local Group confirmed the fact that the method proposed in the present work is simple to use and can provide reliable distance modulus for the galaxies included in our sample. It is highly recommended to reapply the present approach on larger samples having accurate data, particularly the intrinsic apparent magnitudes.

Our revised calculation of $H_0$ using Cepheid in $V$ band calibration leads to a revision of the Cepheids distance for 23 galaxies in the Local Group by a new method, hence calibration of the Hubble constant.

We have used velocities of the studied galaxies as given by Freedman et al. (2001). The velocities are corrected for solar motion with respect to the barycenter of the Local Group as given by Mould et al. (2000). Solar motion solutions lead to slightly larger scattering of Hubble diagram of nearby galaxies. These corrections have been very important to minimize the uncertainty in the measured radial velocities, particularly for the nearest galaxies. For this reason the large-scale distribution of matter in the nearby Universe perturbs the local Hubble flow, causing peculiar motion. To give only one example, the study performed by Willick and Batra (2001) has found values of $H_0 = 85 \pm 5$ km s$^{-1}$Mpc$^{-1}$ and $92 \pm 5$ km s$^{-1}$Mpc$^{-1}$ based on applying different local velocity models to 23 galaxies within 20 Mpc.

Hubble diagram resulting from our calculations to the considered 23 galaxies, using Cepheid distances, is shown in Fig. 3 (B). The values of the constant $H_0$ range between $66 \pm 7$ km s$^{-1}$Mpc$^{-1}$ and $80 \pm 5$ km s$^{-1}$Mpc$^{-1}$. The formal fit to our distance modulus yields a slope of $H_0 = 73 \pm 3$ km s$^{-1}$Mpc$^{-1}$, in a good agreement, to within the uncertainties, with the values of $H_0$ obtained by other methods that extend to much greater distances.

Fig. 4 presents the values of the constant $H_0$ published from 1991 to the present. The recently resulting values of $H_0$ lie in the range of $60-80$ km s$^{-1}$Mpc$^{-1}$. Our result is within this range.

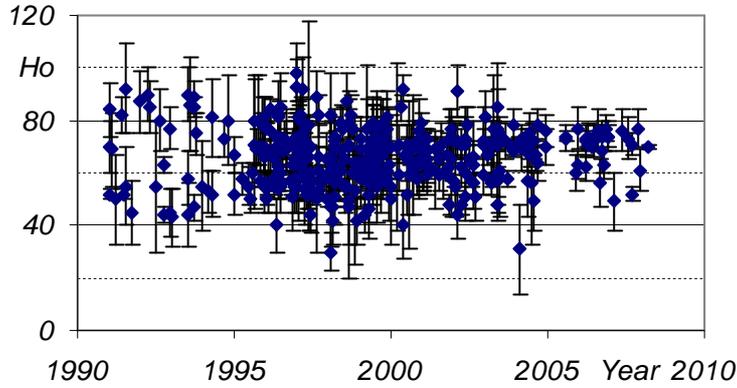

Fig. 4 – Values of the Hubble constant determined from 1990 to the present.

## 8. SUMMARY AND CONCLUSION

We presented an approach in which an assumed Gaussian distribution of a sample of Cepheids in the Local Group is studied. The period and magnitude distribution is studied statistically in 23 galaxies of the Local Group and merge with distance average difference of about $0.2^m$. Also the main sources of uncertainty in this estimate are the reddening correction from the source of observation and from the adopted recession velocity of the cluster. A Gaussian function has been proposed to fit the absolute magnitude and the apparent magnitude of Cepheids in our Galaxy and their period as well. The distance modulus is then the difference between the apparent magnitude of the Extragalactic Cepheids and the absolute magnitude of the Galactic Cepheids at maximum number. From the numerical results, we can provide reliable distance modulus. It would be recommended to reapply the present approach to larger samples with accurate data, in particular for intrinsic apparent magnitude. The number of Cepheids in some galaxies is not enough to make a realistic distribution.

Local variations in the expansion rate due to large-scale velocities make measurement of the true value of $H_0$ difficult. Thus, to get an accurate determination of $H_0$, a large enough volume must be observed to provide a fair sample of the Universe over which an average can be reliable. Almost all the values for $H_0$ lie around the value 73 km s$^{-1}$Mpc$^{-1}$. The scattering in $H_0$ is about 10% and in the majority is caused by errors in the method itself. As an example, we mention the values calculated for the physical parameters, particularly redshift and distance. And, because the values were obtained via different distance methods, based on different physical principles, it is believed that the values are trustworthy. Furthermore, the values for $H_0$ via redshift method and via gravitational lensing are about the same: recent gravitational lensing gives values around $62 \text{ km s}^{-1}\text{Mpc}^{-1}$ (Sandage et al. 2006). This supports the idea that $H_0$ is a real physical constant in the cosmological theories.

*Acknowledgments.* M. Abdel-Sabour, M. S. El-Nawawy, and A. Kordi were supported by KSU, College of Science-Research Center, Project Phys/2008/45.